\documentclass[twocolumn,secnumarabic,amssymb, nobibnotes, aps, prb]{revtex4-1}

\usepackage{graphicx,epsfig,color}
\usepackage{dcolumn}
\usepackage{bm}
\newcommand{\nn}{\nonumber\\}

\newcommand{\bea}{\begin{eqnarray}}
\newcommand{\ea}{\end{eqnarray}}
\newcommand{\eea}{\end{eqnarray}}

\usepackage{graphicx}
\usepackage{amsmath}

\begin{document}
\title{Consistent perturbative treatment of the subohmic spin-boson model yielding arbitrarily small
$T_2/T_1$ decoherence time ratios}
\author{Kyung-Joo Noh and Uwe R. Fischer}
\affiliation{Seoul National University, Department of Physics and Astronomy\\Center for Theoretical Physics,
Seoul 151-747, Korea}
\date{\today}


\begin{abstract}
We present a perturbative treatment of the subohmic spin-boson model which remedies a crucial flaw in previous treatments. The problem is traced back to the incorrect 
application of a Markov type approximation to specific terms in the temporal evolution of the reduced density matrix. 
The modified solution is consistent both with numerical simulations and the exact solution obtained when the bath-coupling spin-space direction is parallel to the qubit energy-basis spin. 
We therefore demonstrate that 
the subohmic spin-boson model is capable of 
describing arbitrarily small 
ratios of the $T_2$ and $T_1$ decoherence times, associated to the decay of the off-diagonal and diagonal reduced density-matrix elements, respectively. An analytical formula for $T_2/T_1$ at the absolute zero
of temperature is provided in the limit of a subohmic bath with vanishing spectral power law exponent. Small ratios closely mimic the experimental results for solid state (flux) qubits, which are subject predominantly to low-frequency electromagnetic noise, and  we suggest a reanalysis of the corresponding experimental data in terms of a {\em nonanalytic} decay of off-diagonal coherence.                   
\end{abstract}
\maketitle


Since the advent of quantum information technology, the problem of two-state systems aka qubits,  
immersed in environmental (bath) degrees of freedom has increasingly gained importance, cf., e.g.,  \cite{Amico,DiVincenzo, vanderWal}. 
The spin-boson model, succinctly describing such a physical situation, namely an open two-state quantum system 
interacting with a bath, accounts for two-state energy relaxation and dephasing,  
and therefore has been studied extensively \cite{Weiss,Leggett}. It was, for example, 
used to describe physical contexts as diverse as flux qubits implemented using superconducting quantum interference devices (SQUIDs) \cite{Makhlin,Governale,Tian,Burkard,Kato}, electron transfer in biomolecules \cite{Garg}, and
phonon coupling in atomic tunneling \cite{Sethna1981}.           

The spin-boson model introduces a continuum of independent simple harmonic oscillators with a given spectral 
weight distribution as the environment, assuming them to couple with the qubit linearly. 
The Hamiltonian of the entire (closed) system therefore reads (setting $\hbar = 1$)
\bea
H&=&H_{0}+V,\quad \mbox{where} \quad H_0 =\frac{1}{2}\omega_{s}\sigma_{z}+\sum_{\bm{k}}{\omega_{\bm{k}}\hat{b}^{\dagger}_{\bm{k}}\hat{b}_{\bm{k}}}, \nonumber\\
V&=&\overrightarrow{n} \cdot \overrightarrow{\sigma} 
\otimes \sum_{\bm{k}}\lambda_{\bm{k}} (\hat{b}_{\bm{k}}+\hat{b}^{\dagger}_{\bm{k}}). \label{spin boson H}
\ea 
Here, $\overrightarrow{\sigma}=(\sigma_x,\sigma_y,\sigma_z)$ are the usual Pauli matrices and $\hat{b}^{\dagger}_{\bm{k}}$,$\hat{b}_{\bm{k}}$ are creation and annihilation operators of harmonic oscillators in the (infinitely extended) bath, labelled by the quantum number(s) $\bm{k}$, which can stand, e.g., for the momentum of the bath excitations. The coupling direction $\overrightarrow{n}$ is parametrized by two angles $\theta$,$\varphi$ as $\overrightarrow{n}=(\sin \theta \cos \varphi ,\sin \theta \sin\varphi,\cos \theta)$.

The bath is conventionally characterized by the quantity $J(\omega) \equiv \sum_{\bm{k}}{\lambda_{\bm{k}}^{2}\delta(\omega-\omega_{\bm{k}})}$, which determines the dynamics of the spin-boson model.
This {\em spectral density} of the bath is effectively a density of states summation weighed by the 
coupling strength squared $\lambda_{\bm{k}}^{2}$, hence the name. 
Usually it is assumed that   
$J(\omega)$ is of a power law form up to a cutoff, $J(\omega) = \eta \omega_{s}^{1-n} \omega^{n}e^{-\frac{\omega}{\omega_{c}}}$, where $\omega_{c}$ is the (physical) cutoff frequency of the bath, assumed to be much larger than the system frequency spacing $\omega_{s}$. The strength of coupling is parametrized by the dimensionless $\eta$. Baths having $n<1, n=1$, and $n>1$ are called subohmic, ohmic, and superohmic, respectively. In a subohmic bath, for decreasing $n$, low-frequency oscillators play an increasingly 
important role. 

The Rabi model, which is a ($\theta=\pi/2$) variant of the spin-boson model with a single-mode ``bath'' (e.g. a two-level atom in a cavity), is integrable \cite{Braak}. 
In principle, an exact solution of the spin-boson model in the same perpendicular coupling case $\theta=\pi/2$ was also obtained \cite{Gardas}, which 
however does not lend itself to a description of the effective qubit dynamics because tracing out the bath is still highly nontrivial. 
Furthermore,  in the simple case that the coupling direction and the energy basis direction of the system are parallel ($\theta = 0$), the spin-boson model is also exactly solvable \cite{Kuang,Nazarov}.

Away from the limiting cases of parallel and perpendicular coupling, $\theta=0,\pi/2$, perturbative approaches have been employed \cite{Makhlin,Governale,Burkard,Nazarov}. 
However, as pointed out in \cite{Nazarov}, the previous perturbative approaches show a pathological behavior for the subohmic heat bath by predicting the instantaneous loss of phase coherence for any finite coupling $\eta$. 
In addition, in the low temperature limit, the perturbative solution does not agree even qualitatively with the exact solution for $\theta = 0$. Previous perturbative approaches thus cannot be self-consistent. 
We aim at remedying this situation, by developing a consistent perturbative approach to the subohmic
spin-boson model which both agrees with the exactly solvable case $\theta=0$ and is
free of the aforementioned pathological behavior. 


The time evolution of the density matrix {of the closed system}  
in the interaction picture, $\tilde{\rho}_{I}(t)$, is determined by the following von Neumann equation 
\begin{equation}
\frac{d \tilde{\rho}_{I}(t)}{dt} = -i[V_{I}(t),\tilde{\rho}_{I}(t)],\label{eq2}
\end{equation} 
where Schr\"odinger and interaction picture operators are related by $A_{I}(t)\equiv U_{H_{0}}^{\dagger}(t) A U_{H_{0}}(t)$. 
We define the reduced density matrix by tracing out the bath, $\rho_{I}(t)\equiv {\rm Tr}_{B}\tilde{\rho}_{I}(t)$,  and employ the following three conditions \cite{Puri,quantiki} 
\begin{itemize}
\item[\{1\}] The Born 
approximation which treats Eq.\eqref{eq2} perturbatively 
to second order accuracy in $V_{I}(t)$.
\item[\{2\}] The initial product state assumption $\tilde{\rho}(0)=\rho(0)\otimes \rho_{B}(0)$ where $\rho_{B}(0)$ is the canonical density matrix at temperature $T$ (or $\tilde{\rho}_{I}(0)=\rho_{I}(0)\otimes \rho_{B}(0)$).
\item[\{3\}] Tr$_{B}[V_{I}(t),\rho_{B}(0)]=0$, 
due to 
$V_I$ having odd and $\rho_{B}(0)$ having even spatial parity. 
\end{itemize}
The equation \eqref{eq2} can then be transformed into the perturbative expression (see
Eqs.\,8.1--8.15 in Ref.\cite{Puri})
\begin{equation}
\frac{d \rho_{I}(t)}{dt} = -\int_{0}^{t}{ds 
{\rm Tr}_{B}[V_{I}(t),[V_{I}(s),\rho_{I}(t)\otimes \rho_{B}(0)]]}. \label{eq3}  
\end{equation}
For the purpose of our derivation to follow, we note that in the literature, one encounters rather widely differing notions of what is called ``Born'' (i.e., weak coupling) 
and ``Markov'' (i.e., short-time memory) approximations, which are not necessarily equivalent. 
For example, Ref.\,\cite{Breuer} derives Eq.\eqref{eq3} by assuming a product state for {\em arbitrary times}  $\tilde{\rho}_I(t) = \rho_I(t) \otimes \rho_{B}(0)$ (instead of only initially as in \{2\}), and terms this ``Born'' approximation. Then the ``Markov'' approximation is imposed that $\rho_I(s)$ in the integrand is replaced by $\rho_I(t)$ to arrive at \eqref{eq3}. On the other hand, \cite{Puri,quantiki} derives Eq.\eqref{eq3} by using the above stated requirements $\{ 1\}$-$\{ 3 \}$, utilizing the Born approximation in the form of 
\{1\}, which can be justified rigorously \cite{Puri}. 
No (variety of) Markov approximation is employed in deriving \eqref{eq3}. 
In addition, the Markov type approximation \{4\} which we will use and explain in detail 
below is {\em distinct} from that used in, e.g., \cite{Breuer,Markovnote}. 

Employing the spin-boson Hamiltonian \eqref{spin boson H} 
in the master equation \eqref{eq3}, {and adopting the variable change $s \rightarrow t-s$}, one obtains 
\begin{multline}
\frac{d \rho_{I}(t)}{dt} = -\int_{0}^{t}{ds}\int_{0}^{\infty}d\omega J(\omega) \\
\times \Big{\lbrace} \coth\left[\frac{\beta\omega}{2}\right] \cos{(\omega s)}
{[\overrightarrow{n} \cdot \overrightarrow{\sigma}(t),[\overrightarrow{n} \cdot \overrightarrow{\sigma}(t-s),\rho_{I}(t)]]}\\
{-i\sin{(\omega s)}[\overrightarrow{n} \cdot \overrightarrow{\sigma}(t),\lbrace \overrightarrow{n} \cdot \overrightarrow{\sigma}(t-s),\rho_{I}(t) \rbrace] \Big{\rbrace}}.
\label{SpinBoson EOM}
\end{multline}
Here, the $t$ dependence of the operators $\overrightarrow{n} \cdot \overrightarrow{\sigma}(t)$ is determined by the spin part of the Hamiltonian, i.e., by $H_{0}^{(s)}=\frac{1}{2}\omega_{s}\sigma_{z}$. 
Previous studies on the spin-boson model \cite{Makhlin,Governale,Burkard,Nazarov}
deduced a solution for the evolution of the 
reduced density matrix $\rho_{I}(t)$ of Eq.\eqref{SpinBoson EOM}, which, when written in the presently
employed notation, reads  
\begin{eqnarray}
\rho_{I,ii}(t) &=& \rho_{I,ii}^{\rm eq}(1-\exp\left[-\frac{t}{T_{1}}\right]) + \rho_{I,ii}(0) \exp\left[-\frac{t}{T_{1}}\right],\; 
\nn
i&=&\{1,2\},
\nonumber\\
\rho_{I,12}(t) &=& \rho_{I,12}(0) \exp\left[-\frac{t}{T_{2}}\right]\exp\left[-i\frac{t}{T_{3}}\right] ,
\nonumber\\
\rho_{I,21}(t) &=& \rho_{I,21}(0) \exp\left[-\frac{t}{T_{2}}\right]\exp\left[i\frac{t}{T_{3}}\right] ,
\label{eq4}
\end{eqnarray}
where the indices refer to qubit space, and $\rho_{I,11}^{eq} =({1+\exp[\beta\omega_{s}]})^{-1}$ and $\rho_{I,22}^{eq} =({1+\exp[-\beta\omega_{s}]})^{-1}$ 
are canonical equilibrium distributions at the inverse temperature $\beta$. Finally, the various rates 
in \eqref{eq4} satisfy  
\bea
\hspace*{-1.5em}\frac{1}{T_{1}} &=& 2\pi \sin^{2}\theta J(\omega_{s})\coth\left[\frac{\beta\omega_{s}}{2}\right], \label{eq5}
\\
\hspace*{-1.5em}\frac{1}{T_{2}} &=& \frac{1}{2T_{1}} + \frac{1}{T_{\phi}}, \quad 
\frac{1}{T_{\phi}} = 4\pi k_{B}T \cos^{2} \theta\, 
\left(\frac{J(x)}{x}\right)_{x \rightarrow 0}  \label{eq7}
\\
\hspace*{-1.5em}\frac{1}{T_{3}} &=&  
2\sin^{2}\!\theta\!\int\limits_{0}^{\infty}\!\int\limits_{0}^{\infty}\!d{s}d\omega J(\omega)  
\coth \left[\frac{\beta \omega}{2}\right]\!
\cos(\omega{s})
\sin(\omega_{s} {s}),\nn \label{eq8}
\ea
where $T_{1}$ and $T_{2}$ are called (energy) relaxation and dephasing {time}, respectively.  
The coupling angle $\varphi$ does not occur in the above: 
The two expressions containing $\varphi$, $(\overrightarrow{n} \cdot \overrightarrow{\sigma})_{12}=\sin\theta e^{-i \varphi}$ and $(\overrightarrow{n} \cdot \overrightarrow{\sigma})_{21}=\sin\theta e^{i \varphi}$ 
enter as absolute squares.

We observe that ${1}/{T_{\phi}}$ in Eq.\eqref{eq7} diverges for the subohmic case $n<1$
at any finite $T$ as $(x^{n-1})_{x \rightarrow 0}$; also note that at absolute zero, $T_{\phi}$ cannot be determined from the formula at all. 
Accordingly, if Eq.\eqref{eq7} is valid for the subohmic regime, this indicates that the dephasing time $T_{2}$ should always vanish. 
This appears to be unphysical, because it implies that phase coherence is destroyed instantaneously as soon as the system comes into contact with a subohmic heat bath, at any value of the coupling strength $\eta$. 
We therefore suspect that Eqs.\eqref{eq4}--\eqref{eq8} do not represent a proper perturbative solution of  
\eqref{SpinBoson EOM}. 

Furthermore, we now argue that Eqs.\eqref{eq4}--\eqref{eq8} are in addition inconsistent with the exact solution for the commuting case $\theta=0$, i.e, when $\overrightarrow{n} \cdot \overrightarrow{\sigma}=\sigma_{z}$. Using a unitary transformation \cite{Kuang}, the exact full density matrix can be found from \eqref{eq2}, only using the initial product state assumption \{2\}. 
The solution for the reduced density matrix, at any coupling strength $\eta$, is then 
\begin{eqnarray}
\rho_{I,11}(t) &=& \rho_{I,11}(0), \quad 
\rho_{I,22}(t) = \rho_{I,22}(0),
\nonumber\\
\rho_{I,ij}(t) &=& \rho_{I,ij}(0)\exp[-Q(t)], \quad i\neq j,
 \label{eq9}\\
\vspace*{1em}
Q(t) &=& 4 \int_{0}^{\infty}d \omega J(\omega)\coth\left[\frac{\beta \omega}{2}\right]
\frac{1-\cos(\omega t)}{\omega^{2}}. \nonumber 
\end{eqnarray}            
To see the inconsistency between Eqs.\eqref{eq4}-\eqref{eq8} (with $\theta=0$) and Eq.\eqref{eq9} in a 
manifest manner, we make Eqs.\eqref{eq4}-\eqref{eq8} concrete for an ohmic heat bath ($n=1$):
\begin{eqnarray}
& \rho_{I,11}(t) = \rho_{I,11}(0), \quad \rho_{I,22}(t) = \rho_{I,22}(0) 
\nonumber\vspace*{0.5em}\\
& \rho_{I,ij}(t) = \rho_{I,ij}(0)\exp[-4\pi\eta k_{B}T t], \quad i\neq j
\label{eq10}
\end{eqnarray} 
In the high temperature limit ($k_{B}T \gg \omega_{c}$), using the approximation $\coth[\frac{\beta \omega}{2}]\simeq \frac{2k_{B}T}{\omega}$ for $\omega$ below the cutoff energy, $Q(t)$ is approximated as $4\pi\eta k_{B}Tt$; hence Eq.\eqref{eq4}-\eqref{eq8} and Eq.\,\eqref{eq9} for large $T$ give the same results and are consistent with each other.    
However, at the absolute zero of temperature, Eq.\,\eqref{eq9} yields the following solution which
obviously does not agree with Eqs.\eqref{eq4}--\eqref{eq8}, i.e., by setting $\coth[\frac{\beta\omega}{2}]=1\, \forall\, \omega$, we get 
\begin{eqnarray}
& \rho_{I,11}(t) = \rho_{I,11}(0), \quad \rho_{I,22}(t) = \rho_{I,22}(0), 
\nonumber\\
& \rho_{I,ij}(t) = \rho_{I,ij}(0)\left(1+(\omega_{c}t)^2\right)^{-2\eta} \quad i \neq j . 
\label{eq11}
\end{eqnarray}
Eq.\eqref{eq10} exhibits an exponential decay of the off-diagonal elements of the reduced density matrix, and the decay rate vanishes at $T=0$. On the other hand, according to the exact solution in Eq.\,\eqref{eq11}, we can still observe a polynomial decay in the off-diagonal elements even at $T=0$. The inconsistency between Eqs.\,\eqref{eq4}--\eqref{eq8} and Eq.\,\eqref{eq9} under the approximations employed so far is therefore apparent.     
We next provide a perturbative solution which avoids the above consistency problems.   


We first observe that the derivation of Eqs.\,\eqref{eq4}-\eqref{eq8} from Eq.\,\eqref{SpinBoson EOM} requires the following two additional conditions to be implemented apart from \{1\}--\{3\}:
\begin{itemize}
\item[\{4\}] The Markov approximation (short-time memory approximation), in the sense of replacing a finite time
integral, $\int_{0}^{t}$, by one extending to {positive infinity}, $\int_{0}^{\infty}$ in Eq.\,\eqref{SpinBoson EOM}. 
\item[\{5\}] $\omega_{s}^{-1} \ll T_{i}\,\,\forall\, i$; that is, averaging out rapidly oscillating terms in Eq.\,\eqref{SpinBoson EOM}. 
\end{itemize} 
We proceed to demonstrate that the 
approximation \{4\} forms the reason 
for the pathological behavior displayed by 
Eqs.\eqref{eq4}--\eqref{eq8}. 
In particular, the problem becomes manifest when $T_{\phi}$ defined in \eqref{eq7} is involved in the solution process (or $T_2$ when $T_1(\theta,\omega_s)$ 
is considered fixed)\cite{Thoss}.

To initiate our discussion, we first observe that one encounters the following four integrals $I_i$ when writing down the right-hand side of 
Eq.\,\eqref{SpinBoson EOM}:    
\begin{eqnarray}
I_{1}+iI_{2}
&=& \int_{0}^{t}{ds \Big{[}\int_{0}^{\infty}{d\omega J(\omega)\coth\left[\frac{\beta\omega}{2}\right]\cos(\omega s) } \Big{]} e^{i\omega_{s}s}},
\nonumber\\
I_{3}
&=& \int_{0}^{t}{ds \Big{[}\int_{0}^{\infty}{d\omega J(\omega)\sin(\omega s) } \Big{]} \sin(\omega_{s}s)}, 
\nonumber
\\
I_{4}
&=& \int_{0}^{t}ds \left[\int_{0}^{\infty}
d\omega J(\omega)\coth\left[\frac{\beta\omega}{2}\right]\cos(\omega s)\right],   \label{Integrals}
\end{eqnarray}
where we used the Euler formula $\exp[i\phi]=\cos\phi +i\sin\phi$ 
to combine the first two integrals into one expression. 
For a spectral density $J(\omega)$ of power law form, the energy integrals over $\omega$ in square brackets yield polynomially decreasing functions of $s$ at $T=0$. The Markov type approximation of replacing $\int_{0}^{t}$ with $\int_{0}^{\infty}$  can then be legitimately applied to $I_1,I_2,I_3$, because these integrals converge sufficiently fast due to rapidly oscillating functions in the integrand such as $\cos(\omega_{s}s)$ and $\sin(\omega_{s}s)$, assuming that condition \{5\} holds. 
In addition, even though $I_1,I_2,I_3$ are functions of both $\omega_{s}$ and $t$, the $\omega_{s}$ dependence 
dominates the $t$ dependence, because the Markovian $t \rightarrow \infty$ limit of these integrals is determined by $\omega_{s}$. On the other hand, due to the absence of such oscillating functions in the integrand, the integral $I_4=I_4(t)$ converges much more slowly. 
Hence it follows that the detailed $t$ dependence of $I_4$ should be treated carefully, and straightforwardly applying the Markov approximation \{4\} to $I_4$ is not warranted.       
As a result, because only $I_4$ is related to the evaluation of $T_{\phi}$, one can obtain a more adequate solution for the reduced density matrix by discarding the Markov approximation $\{4\}$ for
the corresponding ``critical'' terms involving $I_4$.

Applying the Markov type approximation \{4\} 
to $I_{1},I_{2},I_{3}$, using $I_{1}(\infty, \omega_{s}) = \frac{\pi}{2}J(\omega_{s})\coth(\frac{\beta\omega_{s}}{2})$, $I_{3}(\infty, \omega_{s}) = \frac{\pi}{2}J(\omega_{s})$ and noting the definitions for $T_{1},T_{2},T_{3}$ in Eqs.\,\eqref{eq5}-\eqref{eq8}, 
we derive the following differential equations for the reduced density matrix from Eq.\eqref{SpinBoson EOM}, 
\begin{align}
& \partial_{t}\rho_{I,11}(t) = -\pi\sin^{2}\theta J(\omega_{s})\left(\coth\left[\frac{\beta\omega_{s}}{2}\right]+1\right)\rho_{I,11}(t) 
\nonumber\\
&+\pi\sin^{2}\theta J(\omega_{s})\left(\coth\left[\frac{\beta\omega_{s}}{2}\right]-1\right)\rho_{I,22}(t) ,  
\nonumber\\
& \partial_{t}\rho_{I,12}(t) = -\left(\frac{1}{2T_{1}}+\frac{i}{T_{3}}+4\cos^{2}\theta I_{4}(t)\right)\rho_{I,12}(t) ,
\nonumber\\
& \partial_{t}\rho_{I,21}(t) = -\left(\frac{1}{2T_{1}}-\frac{i}{T_{3}}+4\cos^{2}\theta I_{4}(t)\right)\rho_{I,21}(t) ,
\nonumber\\
& \partial_t\rho_{I,22}(t) = -\partial_{t}\rho_{I,11}(t) . \label{1st order ODE for reduced density matrix}
\end{align}
The first and last line can be rewritten in a compactified notation, using 
that $\rho_{I,11}(t)+\rho_{I,22}(t)=1$, 
$\partial_{t}\rho_{I,ii}(t) = -\frac{1}{T_{1}}(\rho_{I,ii}(t)-\rho_{I,ii}^{\rm eq}),\quad i=\{1,2\}$. 
The linear first-order differential equations \eqref{1st order ODE for reduced density matrix} 
are then readily solved by  
\begin{align}
& \rho_{I,11}(t) = \rho_{I,11}^{\rm eq}\left(1-\exp\left[-\frac{t}{T_{1}}\right]\right) + \rho_{I,11}(0) \exp\left[-\frac{t}{T_{1}}\right] ,
\nonumber\\
& \rho_{I,12}(t) = \rho_{I,12}(0)\exp\left[-\frac{t}{2T_{1}}\right]\exp\left[-i\frac{t}{T_{3}}\right]\exp[-K(t)] ,
\nonumber\\
& \rho_{I,21}(t) = \rho_{I,21}(0)\exp\left[-\frac{t}{2T_{1}}\right]\exp\left[i\frac{t}{T_{3}}\right]\exp[-K(t)] ,
\nonumber\\
& \rho_{I,22}(t) = \rho_{I,22}^{\rm eq}\left(1-\exp\left[-\frac{t}{T_{1}}\right]\right) + \rho_{I,22}(0)  \exp\left[-\frac{t}{T_{1}}\right], \label{eq15}
\end{align}
defining a function $K(t)=4\cos^{2}\theta\int_{0}^{t}{dsI_{4}(s)}$, which replaces $t/T_\phi$ in \eqref{eq4},    
\begin{equation}
K(t) = 4\cos^2 \theta \int_{0}^{\infty}{d \omega J(\omega)\coth\left(\frac{\beta \omega}{2}\right)\frac{1-\cos(\omega t)}{\omega^{2}}}. \label{eq16}
\end{equation}
Eq.\,\eqref{eq15} represents an {\em improved} perturbative solution for Eq.\,\eqref{SpinBoson EOM} in that it does not display any of the unphysical pathology exhibited in Eq.\,\eqref{eq7}, and it is reduced to the exact solution
displayed in Eq.\,\eqref{eq9} if we set $\theta=0$. It thus demonstrates perfect consistency with the exact solution for the commuting case. 
Note that with $\theta=0$, $K(t)$ becomes $Q(t)$ in \eqref{eq9},  
and the rates $\frac{1}{T_{1}},\frac{1}{T_{3}}$ vanish. 

\begin{figure}[t]
\centering
\includegraphics[scale=0.275]{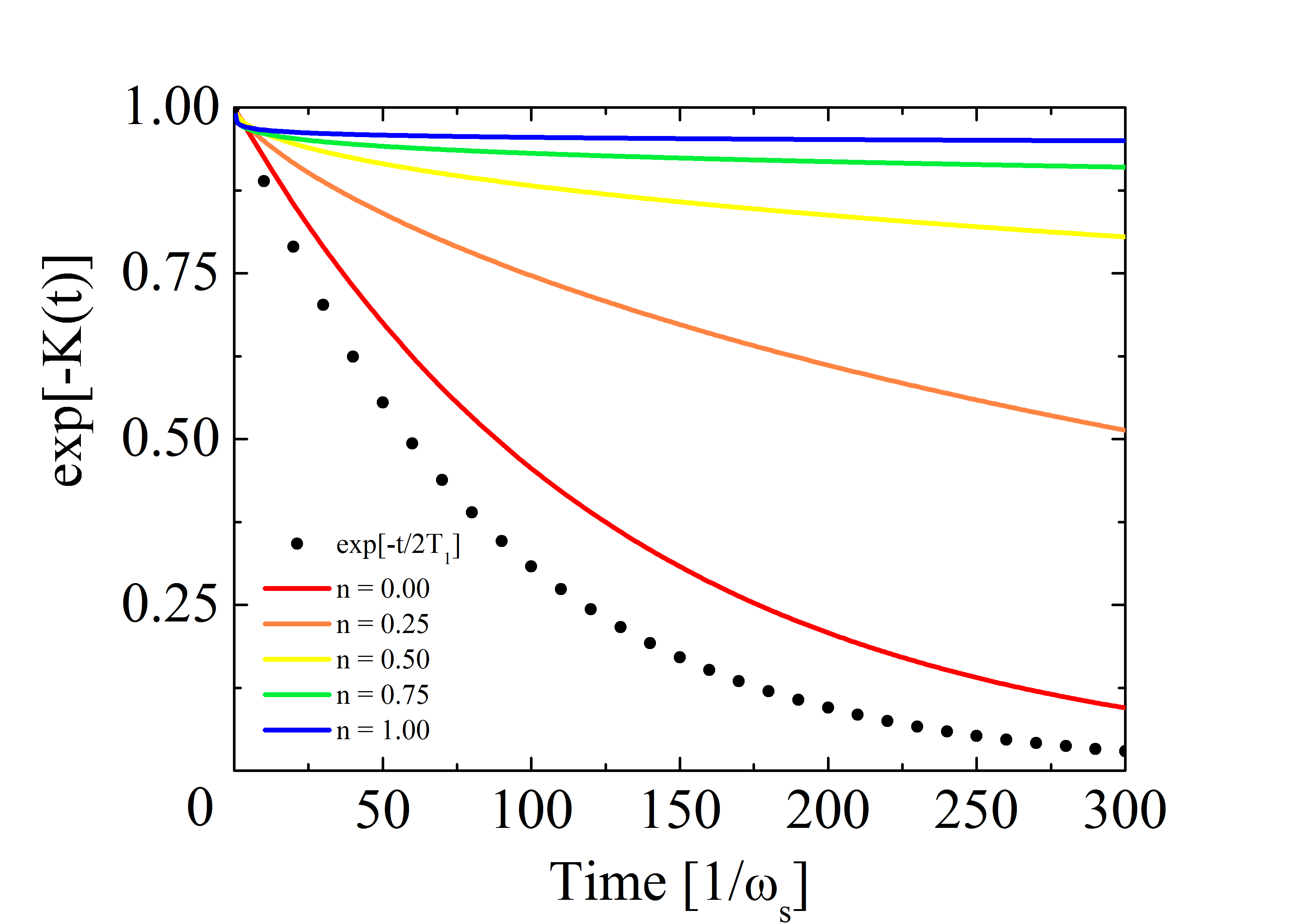}
\caption{The decay factor $\exp[-K(t)]$ in the ohmic and subohmic regime, $0\le n \le 1$; $\eta = 5\times 10^{-3}$ and $\theta = 60\deg$; the ultraviolet cutoff 
$\omega_{c}=100\omega_{s}$. For comparison, the dominant decay factor $\exp[-t/2T_1]$ 
is shown by the dotted line.}
\label{figure 1}
\end{figure}  

At the absolute zero of temperature, by setting $\coth(\frac{\beta \omega}{2}) = 1$ for all $\omega$, $K(t)$ can be calculated for our assumed spectral density of the form $J(\omega) = \eta\omega_{s}^{1-n} \omega^{n}e^{-\frac{\omega}{\omega_{c}}}$. For the subohmic case ($0\le n<1$), $K(t)$ does not diverge even if there is no cutoff energy $\omega_{c}$. In this regime, we may thus let $\omega_{c}\rightarrow \infty$. Then, $K(t)$ is given by $K(t)=2\pi \eta \cos^2 \theta \,\omega_{s} t$ for $n=0$ and $K(t) = 4\eta \cos^2\theta \sin(\frac{n\pi}{2})\frac{\Gamma(n)}{1-n}(\omega_{s} t)^{1-n}$ for $0<n<1$. For the ohmic and superohmic cases ($n\geq 1$), on the other hand, 
$K(t)$ converges only if there is a finite cutoff energy $\omega_{c}$. 
For example, $K(t) = 2\eta \cos^2{\theta} \ln(1+(\omega_{c}t)^2)$ for $n=1$ and $K(t) = 4\eta\cos^2{\theta}\, \frac{\omega_{c}}{\omega_{s}} \frac{(\omega_{c} t)^2}{1+(\omega_{c}t)^2}$ for $n=2$. 
The effect of $K(t)$ is generally strongly cutoff frequency dependent in the superohmic case ($n>1$, not shown in Fig.\,\ref{figure 1}), and increasingly so with spectral power $n$. 
The superohmic case thus needs a more elaborate treatment, which we do not pursue here.


The functions $K(t)$ corresponding to different powers $n$ can be numerically compared with each other, see Fig.\,\ref{figure 1} for the ohmic and subohmic cases. 
In particular, when we approach the zero of temperature, for $n=0$, we obtain an off-diagonal decay which is exactly exponential because then K(t) is linear in t.
The influence of $K(t)$ then becomes comparable to $\exp[-t/2T_1]$ (which dominates for the ohmic regime)--cf.\,Eq.\,\eqref{eq15} and Fig. \ref{figure 1}.

As a result of exponential decay, we can define in this limit of $n\rightarrow 0$ 
the dephasing time $T_{2}$ properly \cite{note n not zero}. Combining Eq.\,\eqref{eq15} and $K(t)=2\pi\eta\cos^2\theta\,\omega_{s} t$ ($n=0$ and $T=0$), the dephasing rate 
$\frac{1}{T_{2}} = \frac{1}{2T_{1}} + 2\pi \eta \omega_{s} \cos^2 \theta  
.$ 
Combining this with Eq.\,\eqref{eq5} 
the ratio $\frac{T_{2}}{T_{1}}$ can be obtained analytically at the absolute zero of temperature for $n\rightarrow 0$, 
\begin{equation}
\frac{T_{2}}{T_{1}} = \frac{2\sin^2 \theta}{2- \sin^2\theta}.\label{eq18}
\end{equation}
Hence, contrary to the ohmic case where the ratio ${T_{2}}/{T_{1}}=2$  regardless of the coupling angle $\theta$, the decoherence time ratio can have any value between $0$ and $2$ depending on the coupling angle $\theta$.
This is of particular relevance for the interpretation of studies on flux qubits created using SQUIDs, see for example  \cite{Vion,Collin,Chiorescu,Yoshihara,Kakuyanagi,Bylander,Stern}, which are typically dominated by low-frequency electromagnetic noise. 
The experiments report a wide range of decoherence time ratios
$\frac{T_2}{T_{1}} \sim 0.01\cdots 2$ rather than this ratio being always identical to two.
Thus we conclude from \eqref{eq18} that the perturbatively treated subohmic heat bath
furnishes a realistic physical description  
for these experimentally realized systems.   
 
\begin{figure}[t]
\centering
\includegraphics[scale=0.275]{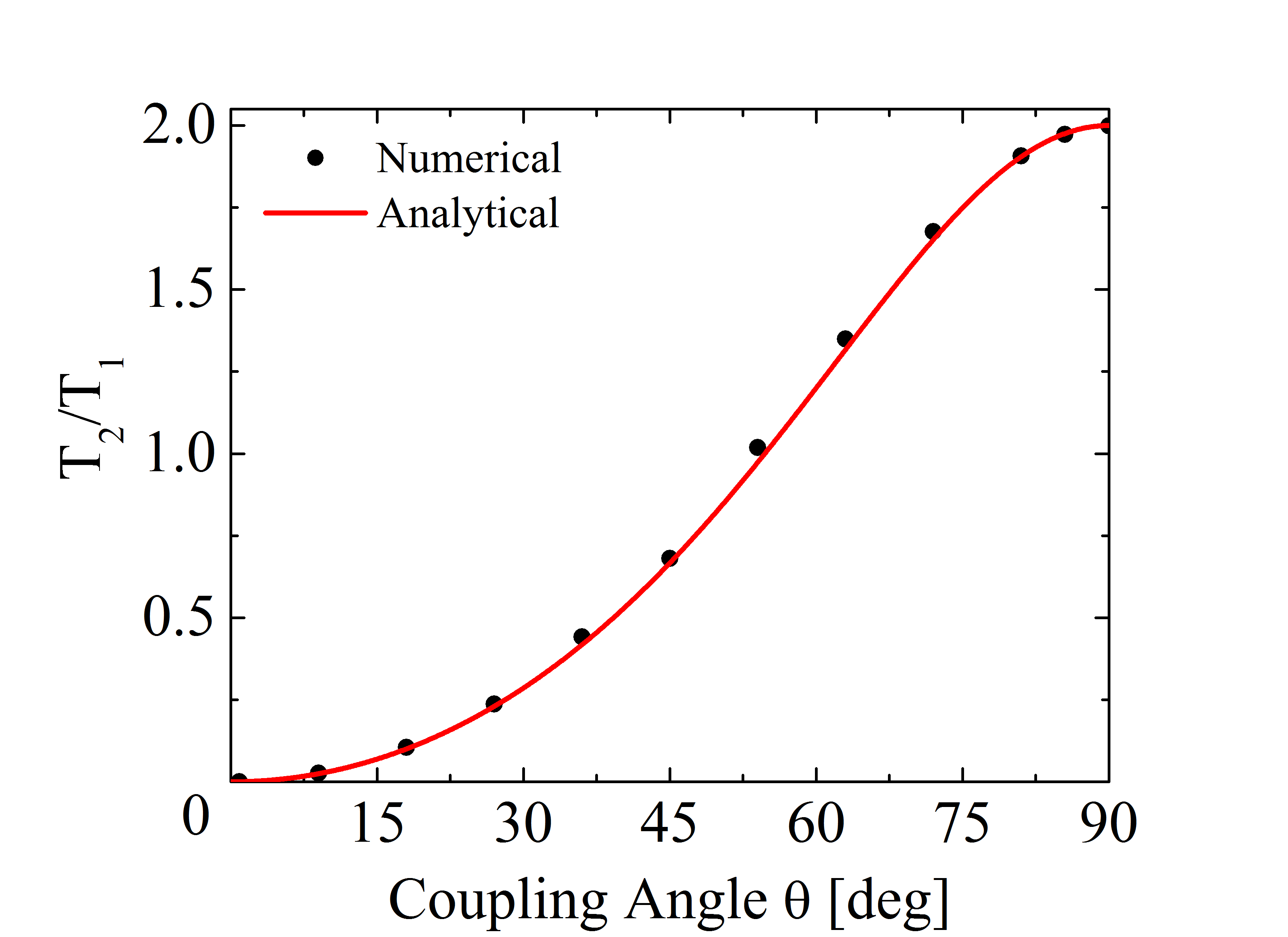}
\caption{Decoherence time ratio $\frac{T_{2}}{T_{1}}$ for the 
extremely subohmic case ($n=0$) as a function of coupling angle $\theta$. 
The analytical result (Eq.\,\eqref{eq18}) is shown by the red line 
and the numerical results by solving Eq.\,\eqref{SpinBoson EOM} are indicated by dots. %
The parameters $\eta$ and $\omega_c$  are the same as in Fig.\ref{figure 1}.
}
\label{figure 2}
\end{figure}

We have validated Eq.\,\eqref{eq18} by checking its consistency with the numerical solution
of \eqref{SpinBoson EOM} for the extremely subohmic case with $\eta = 5\times 10^{-3}$. As can be seen in Fig.\,2, Eq.\,\eqref{eq18} agrees very well with the numerical results (for a smaller coupling $\eta =1 \times 10^{-3}$, numerics and analytics become essentially indistinguishable
on the scale of the figure). On the other hand, with Eqs.\,\eqref{eq4}--\eqref{eq8}, it is readily concluded that it is impossible to get a similarly consistent formula for $T_2/T_1$ 
in the subohmic regime.

To conclude, we have reinvestigated the perturbative solution for the reduced density matrix dynamics
of the subohmic spin-boson model. It was shown that Eqs.\,\eqref{eq15}--\eqref{eq16} represent a solution both consistent with the 
exact solution for 
zero coupling angle, $\theta = 0 $,  
and agrees with the numerical solution of Eq.\,\eqref{SpinBoson EOM} for any coupling direction $\theta$. 
As a result, our solution is capable of describing the physically realistic situation that the dephasing effect of the off-diagonal reduced density matrix elements is much larger than the relaxation effect embodied in the diagonal terms, $T_2\ll T_1$. In addition, the nonlinearity of $K(t)$ we obtained in the subohmic regime $0<n<1$
sheds new light on the interpretation of experiments measuring the off-diagonal decay rates, for example,  Refs.\,\cite{Yoshihara,Kakuyanagi,Stern}. Specifically, we suggest a  (re)analysis of the experimental data 
in terms of a {\em nonanalytic} 
behavior of  
the off-diagonal decay, involving $K(t)\propto t^{1-n}$. We note in this connection that the corresponding formula for $K(t)$, under the conditions explained in the above, in fact also holds for $-1 < n <0$.

{For the parallel case ($\theta=0$), in which there is no energy exchange between spin and heat bath, the  perturbative approach reproduces the exact solution Eq.\,\eqref{eq9}, and hence is valid in the long time limit as well. On the other hand, when there is energy exchange between the spin and the heat bath ($\theta \neq 0$),} nonperturbative studies of the subohmic spin-boson model using the renormalization group flow equation method have pointed out that the perturbative approach for the subohmic case will eventually fail in the long time limit. This is due to the development of a localized, so-called ``trapped'', state for the pseudospin (in a double-well analogy). The phase transition point to the trapped state,  
 $\eta_c=\eta_c(n)$, depends on the power law of the spectral density \cite{Kehrein,Anders,Nalbach,Kirchner}. Despite the potential occurence of a phase transition, the perturbative approach can be applied for the short-time dynamics \cite{Anders,Nalbach}, which in view of the limited coherence times of, in particular, flux qubits will generally suffice. Moreover, a {\em self-consistent} perturbative treatment as presented here {provides an explicit analytic expression for the} reduced density matrix dynamics of the spin-boson model (which the renormalization group approaches did not yield), from which the 
experimentally observed decoherence times can be defined. 

  
This research was supported by the NRF of Korea, Grant No. 2014R1A2A2A01006535. 


\begin{thebibliography}{99}
\bibitem{Markovnote} More specifically, the variety of Markov approximation which is used to derive Eq.3.117 in \cite{Breuer} [and which is equivalent to Eq.\,\eqref{eq3}], from Eq.3.116 therein.
\bibitem{note n not zero} When $0<n<1$, the decay is no longer exponential, due to 
$K(t)\propto t^{1-n}$. 
We have performed an analysis of this somewhat more cumbersome case 
as well, and found the same excellent agreement of numerics and analytics.
\end{thebibliography}
\end{document}